\documentclass[aps,prl,twocolumn,floatfix,longbibliography,superscriptaddress]{revtex4-1}
\usepackage{enumitem}
\usepackage{graphicx}
\usepackage{bm}
\usepackage{epsf}
\usepackage{ulem}
\usepackage[colorlinks = true]{hyperref}
\usepackage{amssymb}
\usepackage{amsmath}
\usepackage{graphicx}
\usepackage{rotating}
\usepackage{epsfig}
\usepackage{psfrag}
\usepackage{amsmath}
\usepackage{subfigure}

\newcommand{\gtappr}{{{\lower4pt\hbox{$>$} } \atop \widetilde{ \ \ \ }}}

\newlength{\figwidth}
\newcommand{\fg}[3]
{\begin{figure}[htb]\vspace*{-0cm}\centerline{\includegraphics[width=\figwidth]{#1}}\vskip
-0.2cm \caption{\label{#2}#3}\end{figure}}

\begin{document}
\title{Quantum Annealed Criticality}
\author{Premala Chandra}
\affiliation{Center for Materials Theory, Rutgers University, Piscataway, New Jersey, 08854, USA} 
\author{Piers Coleman}
\affiliation{Center for Materials Theory, Rutgers University, Piscataway, New Jersey, 08854, USA} 
\affiliation{Department of Physics, Royal Holloway, University of London, Egham, Surrey TW20 0EX, UK}
\author{Mucio A. Continentino}
\affiliation{Centro Brasileiro de Pesquisas Fiscicas, 22290-180,
	Rio de Janeiro, RJ, Brazil}
\author{Gilbert G. Lonzarich}
\affiliation{Cavendish Laboratory, Cambridge University, Cambridge CB3 0HE, United Kingdom} 

\date{\today}

\begin{abstract}
Experimentally there exist many materials with first-order phase transitions
at finite temperature that display quantum criticality.  Classically, a
strain-energy density coupling is known to drive first-order
transitions in compressible systems, and here we generalize this Larkin-Pikin\cite{Larkin69b}
mechanism to the quantum case. We show that if the $T = 0$ system lies above
its upper critical dimension, the line of first-order transitions can end in a
“quantum annealed critical point” where zero-point fluctuations restore the
underlying criticality of the order parameter.
\end{abstract}
\maketitle

The interplay of first-order phase transitions with quantum 
fluctuations 
is an active area 
\cite{Belitz99,Grigera01,Chubukov04,Belitz05,Maslov06,Rech06,Kirkpatrick12,Brando16} in the study of 
exotic quantum states near zero-temperature phase transitions
\cite{Chandra90,Chandra95b,Balents10,Norman16,Fernandes18}. 
In many metallic quantum ferromagnets, coupling of the magnetization to 
low energy particle-hole excitations transforms a high temperature 
continuous phase transition  into a low temperature
discontinuous one, and 
the resulting classical
tricritical points have been observed in many systems
\cite{Belitz99,Grigera01,Chubukov04,Belitz05,Maslov06,Rech06,Kirkpatrick12,Brando16}. 
Experimentally there also exist insulating materials that have 
classical first-order transitions that display quantum criticality  
\cite{Ishidate97,Suski83,Horiuchi15,Rowley14,Chandra17},  
and here we provide a theoretical basis for this observed behavior.

At a first-order transition the quartic mode-mode coupling of 
the effective action becomes negative.
One mechanism for this phenomenon, studied by
Larkin and Pikin \cite{Larkin69b} (LP),
involves the interaction of 
strain with a fluctuating critical order parameter. 
LP found that a diverging specific 
heat in the clamped 
system of fixed dimensions leads to a first-order 
transition in the unclamped system at constant pressure. 
Specifically, the Larkin-Pikin criterion \cite{Larkin69b,Bergman76} for
a first order phase transition is
\begin{equation}\label{LP}
\kappa \ <  \ \frac{\Delta C_{V}}{T_{c}} \left(\frac{dT_{c}}{d \ln V} \right)^{2}
\end{equation}
where $V$ is the volume, $\Delta C_{V}$ is the singular 
part of the specific heat capacity in the clamped system, 
$T_{c}$ is the transition temperature
and $\frac{dT_{c}}{d {\rm ln} V}$ is its strain derivative. 
The effective bulk modulus $\kappa$ is defined
as $\kappa^{-1} = {\it K}^{-1} - ({\it K} + \frac{4}{3} \mu)^{-1}$
where 
${\it K}$ and $\mu$ are the bare bulk and the 
shear moduli in the absence of coupling to the order parameter fields;  
more physically $\kappa \sim {\it K} \frac{c_L^2}{c_T^2}$ where
$c_L$ and $c_T$ are the longitudinal and the transverse sound 
velocities \cite{Landau86}.  We note that shear strain plays a crucial
role in this approach that requires $\mu > 0$.
Short-range fluctuations in the atomic displacements renormalize the 
quartic coupling of the critical modes, but it is the coupling of the
uniform ($q=0$) strain to the energy density, the modulus squared of the 
critical order parameter, that results in a
{\sl macroscopic} instability of 
the critical point leading to a discontinuous transition.

\figwidth=\columnwidth
\fg{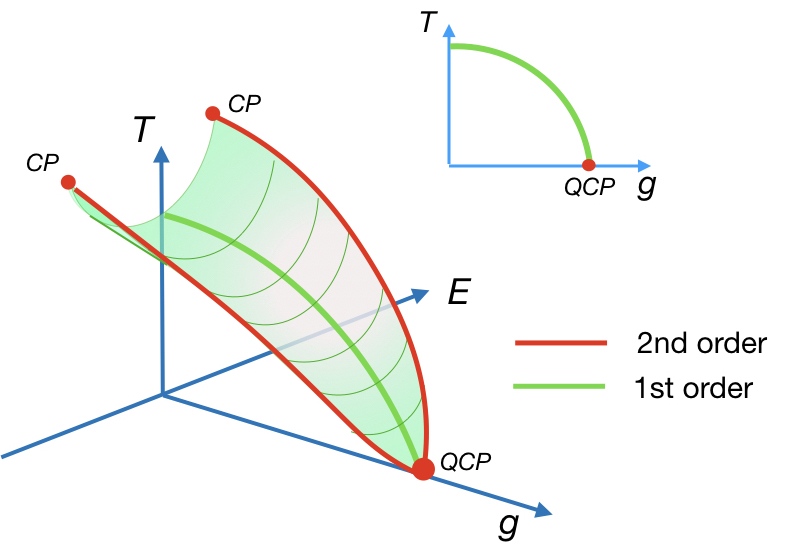}{fig1}{Proposed Temperature-Field-Pressure Phase Diagram
with a sheet of first-order transitions bounded by second-order phase lines linking the three critical points, two classical and one quantum;  
Inset:  Temperature-Pressure ``slice'' indicating a line of classical phase transitions ending in a ``quantum annealed critical point'' with the 
standard temperature fan where the underlying order parameter criticality is
restored by zero-point fluctuations.}

Here we rewrite the Larkin-Pikin criterion in terms of 
correlation functions so that it can be generalized to the quantum
case. We show that if the $T=0$ quantum system lies above its 
upper critical dimension, 
the corrections to the renormalized bulk modulus are non-universal; 
the line of classical first-order transitions 
can end in a ``quantum annealed critical point'' where
zero-point fluctuations restore the underlying 
criticality of the order parameter.
We end with a discussion of the temperature-field-pressure phase diagram and
specific measurements to probe it (cf. Fig. \ref{fig1}).

Low-temperature measurements on ferroelectric insulators 
provide a 
key motivation for 
our study \cite{Ishidate97,Suski83,Horiuchi15,Rowley14,Chandra17}.
At finite temperatures and ambient pressure these materials often display 
first-order transitions due to strong electromechanical 
coupling \cite{Lines77};
yet in many cases \cite{Ishidate97,Suski83,Horiuchi15,Rowley14,Chandra17} 
their dielectric susceptibilities
suggest the presence of pressure-induced quantum criticality associated with 
zero-temperature continuous
transitions \cite{Ishidate97,Suski83,Horiuchi15,Rowley14,Chandra17}. 
It is thus natural to explore whether a quantum generalization of the
Larkin-Pikin approach \cite{Larkin69b}, involving the coupling of critical 
order parameter 
fluctuations to long wavelength elastic degrees of freedom, can be 
developed to describe this phenomenon.

In the simplest case of a scalar order parameter $\psi$ and isotropic
elasticity, the Larkin-Pikin (LP) mechanism \cite{Larkin69} refers to a system where the
order parameter $\psi (\vec{x})$ is coupled to the volumetric
strain with interaction energy
\begin{equation}\label{}
H_{I}  = \lambda \int d^{3}x \ e_{ll} (\vec{x}) \ \psi^{2} (\vec{x})
\end{equation}
where $e_{ab} (\vec{x}) = \frac{1}{2}\left(
\frac{\partial u_{a}}{\partial x_{b}}
+\frac{\partial u_{b}}{\partial x_{a}}
 \right)$ is the strain tensor, $u_{a} (\vec{x})$ is the atomic displacement,
 $e_{ll} (x)={\rm
 Tr}[e(\vec{x})]$ is the volumetric strain and $\lambda$ is a
coupling constant associated with the strain-dependence of $T_c$,
$\lambda = \left(\frac{d T_c}{d {\rm ln} V}\right)$. 
Though the elastic degrees of freedom are assumed to be Gaussian, and
thus  can be formally integrated out exactly, this must be done with
some care.   
This is because the strain field separates
into a uniform ($\vec{q}=0$) term defined by boundary conditions and 
a finite-momentum ($\vec{q} \neq 0$) contribution determined by fluctuating atomic
displacements
\begin{equation}\label{splity}
e_{ab} (\vec{x}) = e_{ab} + \frac{1}{V}\sum_{\vec q  \neq  0}
\frac{i}{2}[ q_{a}u_{b} (\vec{q})+q_{b}u_{a} (\vec{q})]e^{i\vec{q} \cdot \vec{x} },
\end{equation}
where
$\{a,b\} \in [1,3]$ and
$u_{a} (q)$ is the Fourier transform of $u_a(x)$. 
Here we employ periodic boundary conditions to a finite size system
with volume $V=L^{3}$ and discrete momenta $\vec{q}= \frac{2\pi}{L}
(l,m,n)$, where $l, m, n$  are integers.

The uniform strain vanishes when the crystal is
externally clamped. 
 The main effect of
integrating out the finite wavevector fluctuations in the strain is 
to induce a finite correction to the
short-range interactions of the critical
fluctuations that can be absorbed into the quartic $\psi^{4}$ terms in
the action. 
By contrast, fluctuations in the uniform
component of the strain induce an infinite-range attractive interaction between
the critical modes (see Supplementary Materials), and it is this
component of the interactions that is responsible for driving first
order behavior. 
The problem is then reduced to the interaction of
critical order parameter modes, mediated by the fluctuations 
of a uniform
strain field $\phi$ 
with bulk modulus
$\kappa$ (for details 
see Supplementary Materials).
Conceptually, 
the Larkin-Pikin approach amounts to a  study of critical phenomena in
a clamped system, followed by a stability analysis of the
critical point once the clamping is removed.

Recently it was proposed to adapt the Larkin-Pikin approach to 
pressure($\cal{P}$)-tuned quantum magnets
where it is often found
that $\frac{dT_c}{d\cal{P}} \rightarrow \infty$ as $T_c \rightarrow 0$; 
the authors argued that the associated quantum phase 
transitions should then be first-order \cite{Gehring08,Gehring10,Mineev11}.   
However such a diverging coupling of the 
critical order parameter fluctuations and the lattice should lead to 
structural instabilities near the quantum phase transition
that have not been observed \cite{Brando16,Bean62}. 
Furthermore dynamics must be included when treating 
thermodynamic quantities at zero temperature \cite{Sachdev99,Continentino17}. 

We recast the Larkin-Pikin
criterion in the language of correlation functions,  
generalizing the LP approach to the quantum case summing
over all possible space-time configurations.
The strain field again separates
into two contributions as in equation (\ref{splity}), one associated with static uniform boundary
conditions and the other determined by short wavelength 
displacements fluctuating at all frequencies   
\begin{equation}\label{splity2}
e_{ab} (\vec{x},\tau) = e_{ab} + \frac{1}{\beta V}
\sum_{i\nu_n}
\sum_{\vec q  \neq  0}
\frac{i}{2}[ q_{a}u_{b} (q)+q_{b}u_{a} (q)]e^{i(\vec{q} \cdot \vec{x} - \nu_n\tau)}
\end{equation}
where  
$q_\alpha \equiv (\vec{q},i\nu_n)$ 
with $\alpha \in [1,4]$, $u_b(q) \equiv u_b(\vec{q},i\nu_n)$ and
$\nu_n= 2 \pi n T$ is a Matsubara frequency ($k_{B}=1$). A detailed 
analysis indicates 
that when these space-time elastic degrees of freedom are integrated out,
they lead to the coupling of the quantum critical order parameter modes
to a {\sl classical} strain field $\phi$, uniform in both space and
time, with the same effective 
bulk modulus $\kappa$ as in the finite-temperature case
 (see Supplementary Material).
The resulting effective action takes the form
\begin{align}\label{action}
S_{eff}[\psi,\phi ] = \int_{0}^{\beta }d\tau
\int d^{3}x 
  \left[{\cal{L}}[\psi]+ \lambda \phi \ \psi^{2} (\vec{x},\tau ) +
\frac{1}{2}\kappa \phi^{2}\right],
\end{align}
where $ (\vec{x},\tau)$ are the Euclidean space-time
co-ordinates and $\cal{L[\psi]}$ is the Lagrangian of the order 
parameter $\psi(\vec{x},\tau)$ that undergoes a continuous transition in
the clamped system;
in the simplest case $\cal{L[\psi]}$ is a $\psi^{4}$ field
theory
\begin{equation}\label{l3}
{\cal L}[\psi] = 
\frac{1}{2} (\partial_{\mu}\psi
)^{2}+ \frac{a}{2}\psi^{2}+ \frac{b}{4!}\psi^{4}.
\end{equation}

\figwidth=\columnwidth \fg{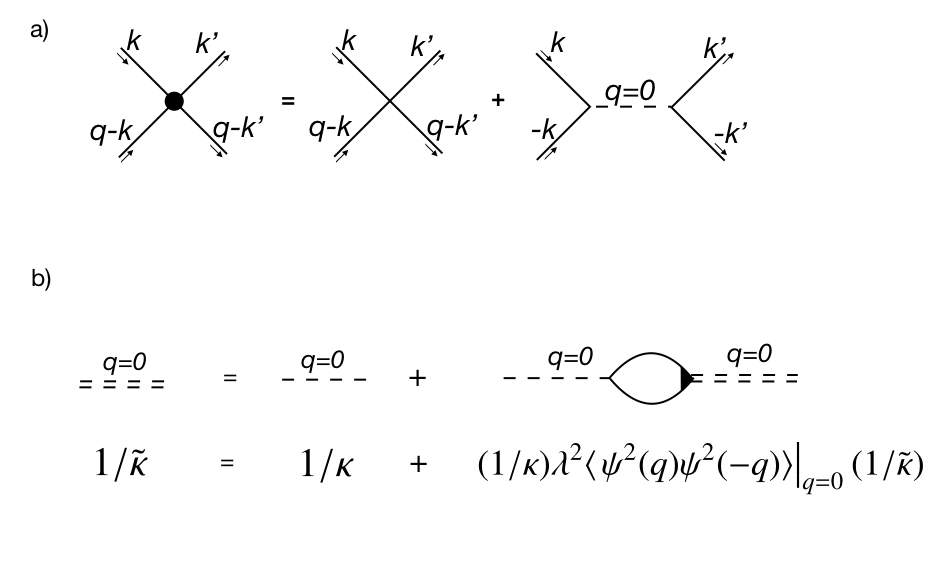}{fig2}
{Diagrammatic approach to the generalized Larkin-Pikin criterion a)
Bare interaction is a sum of a local and a nonlocal contribution
mediated by fluctuations in the strain; b) Feynman diagram showing
renormalization of the strain propagator by coupling to energy fluctuations.
}

The partition function of the unclamped system is then 
\begin{equation}\label{}
Z[\phi ] = e^{-\beta F[\phi ]}=\int {\cal D}[\psi]e^{-S_{eff}[\psi,\phi ]}, 
\end{equation}
where the trace is over the internal variable $\psi$, and $Z[\phi ]$
to be evaluated at the stationary point $F'[\phi ]=0$.
The renormalized bulk modulus, $\tilde{\kappa} = \kappa - \Delta \kappa$, is
\begin{equation}\label{effkap}
\tilde \kappa = \frac{1}{V}\frac{\partial^{2}F}{\partial \phi^{2}} = \kappa -
\lambda^{2}
\int d^{3}xd\tau \langle \delta \psi^{2} (\vec{x},\tau )\delta  \psi^{2} (0)\rangle, 
\end{equation}
where $\delta \psi^{2} (\vec{x},\tau )= \psi^{2} (\vec{x},\tau )- \langle  \psi^{2}\rangle$. 
In the classical problem there is no time-dependence, and 
$\int_{0}^{\beta } d\tau
\rightarrow \beta \equiv 1/ T$ 
so at the transition
\begin{equation}\label{effkapclas}
\tilde{\kappa }= 
\frac{1}{V}\frac{\partial^{2}F}{\partial \phi^{2}} = \kappa -
\frac{\lambda^{2}} {T_c}
\int d^{3}x\langle \delta \psi^{2} (\vec{x} )\delta  \psi^{2} (0)\rangle =  \kappa - \Delta \kappa. 
\end{equation}
$\Delta \kappa$ 
in (\ref{effkapclas}) 
is proportional to 
energy fluctuations, and can be re-expressed
as $\frac{\lambda^2}{T_c} {\Delta C_V}$; 
we thus recover the LP criterion (\ref{LP}) 
($\kappa < \Delta \kappa$ or $\tilde{\kappa}< 0$) for
a first-order transition.
We note that the  renormalized quartic mode-mode coupling coefficient
associated with (\ref{action}) 
changes sign concomitantly with the renormalized bulk modulus; the former 
has 
contributions from both the strain 
coupling and from higher order parameter fluctuations \cite{Das13}.

The renormalized bulk modulus $\tilde{\kappa}$
can also be obtained diagrammatically (cf. Figure \ref{fig2}).  
In the low-energy effective action, the quartic term now has a contribution 
from the coupling of the order parameter fluctuations to the effective 
uniform strain.  We then can use a Dyson equation for the 
strain propagator to determine $\tilde{\kappa}$.
More specifically we can write
\begin{equation}\label{}
\left(\frac{1}{{\tilde \kappa} }\right) = \left(\frac{1}{\kappa }\right) + 
\left(\frac{1}{\kappa }\right) \lambda^2  \langle \psi^2(q) 
\psi^2(-q) \rangle \big\vert_{q=0} \left(\frac{1}{{\tilde \kappa} }
\right)
\end{equation}
%that results in 
%\begin{equation}\label{LPresponse}
%{\tilde \kappa} = \kappa - \Delta \kappa = \kappa - \lambda^2 
%\langle \psi^2(q) \psi^2 (-q)\rangle \vert_{q=0}.
%\end{equation}
that results in  
\begin{equation}\label{kappaeff}
\tilde{\kappa}= \kappa - \Delta \kappa = \kappa - \lambda^{2}\chi_{\psi^{2}} 
\end{equation}
where $\chi_{\psi^{2}}= \left.
\chi_{\psi^2} (\vec q,i\nu_n) \right\vert_{\vec{ q},i\nu_n=0}$ is the
static susceptibility for $\psi^2$, where
\begin{equation}\label{chip2}
\chi_{\psi^{2}} (\vec{q},i\nu_n )=\int_{0}^{\beta } d\tau \int 
{d^{d}x} \langle \delta \psi^{2} (\vec{x},\tau) \delta \psi^{2} (0)\rangle 
e^{i \nu_n \tau - i \vec{q}\cdot \vec{x}},
\end{equation}
is the Fourier transform of the fluctuations in $\psi^{2}$ and
$d=3$.
The sign of $\tilde \kappa$ in (\ref{kappaeff})
is determined by the infrared 
behavior of $\Delta \kappa$; if it diverges, as it does classically (for 
a scalar order parameter and isotropic elasticity), 
then this correction is universal and the transition is first order.

Another possibility 
is revealed in the zero-temperature
long-wavelength Gaussian approximation of (\ref{kappaeff}). If we make
the Gaussian approximation $\langle \delta\psi^{2} (x)\delta \psi^{2}
(0)\rangle \approx (\langle \delta \psi (x)\delta \psi (0)\rangle )^{2}$, then 
\begin{equation}
\lim_{T \rightarrow 0} \Delta \kappa \propto \int dq \  d\nu \ q^{d-1} \ [\chi_{\psi}(\vec{q},i\nu)]^{2}
\end{equation}
where $\chi_{\psi } (\vec{q},i\nu)$, the order parameter
susceptibility, is the Fourier transform of the correlator $\langle
\psi (x)\psi (0)\rangle $. 
Since dimensionally $[\chi] = \left[\frac{1}{q^2}\right]$ and $[\nu]=[q^z]$,
we find that in the approach to the quantum phase transition 
\begin{equation}\label{dimGauss}
\lim_{T \rightarrow 0} \ [\Delta \kappa] = \frac{[q^{d+z}]}{[q^4]} 
\end{equation}
so that the quantum corrections to $\kappa$ are non-singular for $d+z>4$. The
presence of quantum zero-point fluctuations increases the effective dimensionality of the phase space for order parameter fluctuations.  If the effective dimensionality of the quantum system lies above its upper critical dimensionality, this will have the effect of liberating the quantum critical point from the inevitable infrared slavery experienced by its finite-temperature classical 
counterpart.  
In particular the correction to the renormalized bulk modulus is 
then non-universal, allowing for quantum annealed criticality
where zero-point fluctuations toughen the system against the macroscopic
instability present classically, restoring its underlying continuous 
phase transition. 

We have therefore identified a theoretical scenario where there is a quantum continuous transition even though all transitions at finite temperature are 
first-order.  Application of a field conjugate and parallel/antiparallel to the order parameter in such a system leads to a line of first-order transitions ending in two classical critical points.  Therefore by continuity there is a surface of first-order phase transitions in the 
phase diagram (cf. Figure \ref{fig1}) connecting the three critical points, one quantum and two classical, bounded by second-order phase lines. This phase diagram then presents an alternative scenario of the interplay of discontinuous transitions and fluctuations to that studied in metallic magnets where applied field is needed to observe quantum criticality in addition to the tuning parameter \cite{Brando16}.

The specific heat exponent $\alpha$ plays a key role in the universality of the classical Larkin-Pikin criterion (\ref{LP}) since 
the coupling of the order parameter to the lattice is a strain-energy density. For the scalar ($n=1$) case considered 
here, $\alpha > 0$, so that $\Delta \kappa$ is singular and the finite-temperature transition is always first-order; for 
$d + z > 4$, there is a quantum annealed criticality but no quantum 
tricritical point since the quartic mode-mode term in 
the effective action jumps from negative to positive due to the 
change of effective dimension. 

For systems with multi-component order parameters ($n \ge 2$), $\alpha$ is 
negative so the correction to the renormalized bulk modulus 
will be nonuniversal even at 
finite temperatures \cite{Bergman76,Bruno80,deMoura76}.  
In this case, there can be a classical tricritical point at finite pressures
with a second-order transition that continues to zero temperature;
this situation should be robust to everpresent disorder following
the Harris criterion \cite{Harris74}. 
By contrast everpresent elastic anisotropy is known to 
destabilize criticality in the classical isotropic elastic scalar ($n=1$) 
lattice and to drive it first-order 
into an inhomogeneous state \cite{Bergman76,Bruno80,deMoura76};  
here quantum annealed criticality may still  be possible 
due to the increase of effective dimensionality.  The coupling of domain
dynamics to anistropic strain has been studied 
classically for ferroelectrics \cite{Brierley14}, and implications for
the quantum case are a topic for future work.

Because of its underlying non-universal nature, 
the possibility of pressure-tuned quantum annealed criticality must 
be determined in specific setttings.
Ferroelectrics have a dynamical exponent $z=1$, so such three-dimensional
materials are in their 
marginal dimension;
logarithmic corrections to the bulk modulus are certainly 
present but they are not 
expected to be singular.
Indeed such contributions to the dielectric 
susceptibility, $\chi$, 
in the approach to ferroelectric 
quantum critical points have not been observed to 
date \cite{Rowley14}; furthermore here the temperature-dependence of 
$\chi$ is described well by a self-consistent 
Gaussian approach appropriate above its upper critical dimension \cite{Rowley14,Chandra17}. Therefore there may be a very
weak first-order quantum phase transition but experimentally 
it appears to be indistinguishable from 
a continuous one.  We note that near quantum criticality the main effect of 
long-range dipolar interactions, not included in this treatment, 
is to produce a gap in the logitudinal fluctuations, but the 
transverse fluctuations remain critical \cite{Rechester71,Khmelnitskii73,Roussev03}; the excellent agreement between theory and experiment at ferroelectric quantum criticality confirms that this is the case \cite{Rowley14,Chandra17}.  

Dielectric loss and hysteresis measurements can be 
used to probe the line of classical 
first-order transitions, and to determine the nature of the 
quantum phase transition.  The Gruneisen ratio ($\Gamma$), the 
ratio of the thermal expansion and the specific heat, is known to change signs across the quantum phase 
transition \cite{Zhu03,Garst05}; furthermore it is predicted to diverge at a 
3D ferroelectric quantum critical point as $\Gamma \propto \frac{1}{T^2}$ 
so this would be a good indicator 
of underlying quantum criticality \cite{Chandra17}.
Both the bulk modulus and the longitudinal sound velocity should display jumps 
near quantum annealed criticality, though specifics are 
material-dependent since the fluctutation contributions to both are 
non-universal.

In summary, we have developed a theoretical framework to describe compressible 
insulating systems that have classical first-order transitions 
and display pressure-induced quantum criticality.  We have generalized the Larkin-Pikin criterion \cite{Larkin69b} in the language of correlation and response functions; from this standpoint it is clear that the correction to the renormalized bulk modulus, singular at finite temperature, is non-universal at $T=0$ for $d+z > 4$  so then the quantum transition may be continuous.  
Our analyis has been performed for the case of a scalar order parameter and isotropic elasticity where the phase transition is first-order for all finite temperature; in this extreme instance we argue that it is still possible to have quantum annealed criticality. Naturally the presence of a finite-pressure classical tricritical point ensures a continuous quantum phase transition.  The key point is that a compressible material can host a quantum critical phase even if it displays a first-order transition at ambient pressure.  More generally the order of the classical phase transition can be different from its quantum counterpart. 

We note in ending that there are experiments on metallic 
systems \cite{Schmiedeshoff11,Tokawa13,Steppke13} that also
suggest quantum annealed criticality, so a quantum
generalization of the electronic case \cite{Pikin70}  
with possible links to previous work on metallic magnets should be 
pursued \cite{Brando16}; implications for doped paraelectric materials and polar metals \cite{Chandra17} will also be explored. Extension of this work to
quantum transitions between two distinct ordered states separated by first-order classical transitions may be relevant to the iron-based
superconductors \cite{Quader14} and to the enigmatic heavy 
fermion material 
$URu_2Si_2$ where quantum critical endpoints have been suggested
\cite{Chandra13}.
Finally the possibility of quantum annealed criticality in 
compressible materials, 
magnetic and ferroelectric, provides new settings for the exploration
of exotic quantum phases where a broad temperature range 
can be probed with easily accessible pressures
due to the lattice-sensitivity of these systems. 

We have benefitted from discussions with colleagues including
A.V. Balatsky, L.B. Ioffe, D. Khmelnitskii and P.B. Littlewood. PC
and PC  gratefully acknowledge the Centro Brasileiro de Pesquisas 
Fisicas (CBPF), Trinity College (Cambridge) and the Cavendish Laboratory
where this project was initiated.  
Early stages of this work were supported by 
CAPES and FAPERJ grants CAPES-AUXPE-EAE-705/2013 (PC and PC), 
FAPERJ-E-26/110.030/2103 (PC and PC), and
NSF grants DMR-1309929 (P. Coleman)
and DMR-1334428 (P. Chandra).
MAC acknowledges the Brazilian agencies CNPq and FAPERJ for partial
financial support.
GGL acknowledges support from grant no. EP/K012984/1 of the ERPSRC and the CNPq/Science without Borders Program.
PC, PC and GGL thank the Aspen Center for Physics and NSF grant PHYS-1066293 
for hospitality where this work was further developed and discussed.
PC and PC thank S. Nakatsuji and the Institute for Solid State Physics (U. Tokyo) for hospitality where this work was completed.

%-------------------
%\begin{thebibliography}QACP}
%-------------------
%\end{thebibliography}

\bibliography{QACP} %bibliography file with bibtex
%-------------------------------------------------------------------------------

\onecolumngrid
\newpage

\section{Supplementary Material  for Quantum Annealed Criticality}
\subsection{Overview}\label{}

The key idea of the Larkin-Pikin approach is that we integrate out the Gaussian
strain degrees of freedom from the action to derive an effective
action for the order parameter field so that 
\begin{equation}\label{summary}
Z = \int {\cal D}[u ]
\int {\cal D}[\psi]e^{- S[\psi,u]} \quad \quad  
\longrightarrow 
\quad \quad Z = \int {\cal D}[\phi ]
\int {\cal D}[\psi]e^{- S_{eff}[\psi ,\phi]}.  
\end{equation}
The key element in this
procedure is a separation of the strain field into uniform and 
fluctuating components. When we integrate out the uniform component of
the strain, it induces an infinite-range attractive interaction between the
order parameter modes mediated by a {\sl classical}  
field $\phi$ that is uniform in both space and time. 
The main effect of the integration of the 
fluctuating strain component is to renormalize the short-range
interactions between the order parameter modes; however completion of
the Gaussian integral also leads to an infinite range repulsive order
parameter interaction. The overall infinite range interaction
is attractive, but this subtlety needs to be checked carefully in both
the classical and quantum cases, as is performed explicitly in this
Supplementary Material; here we summarize its main results.  The relevant
quantum generalization of the effective action in (\ref{summary}) is
\begin{equation}\label{qSeff}
S_{eff}[\psi,\phi] = \int d^{4}x
\left[\frac{\kappa}{2}\phi^{2}+ \frac{P^{2}}{2K}+ {\cal L}[\psi,b^{*} ] + \lambda \left(\phi+ \frac{P}{K}\right) \psi^{2}[x] \right]
\end{equation} 
where
\[
{\cal L}[\psi,b^{*} ]= 
\frac{1}{2} (\partial_{\mu}\psi
)^{2}+ \frac{a}{2}\psi^{2}+ \frac{b^{*}}{4!}\psi^{4}
\]
is the $\psi^{4}$ Lagrangian, with a renormalized short-range
interaction
\begin{equation}\label{bstar}
b^{*}= b - \frac{12\lambda^{2}}{K+\frac{4}{3}\mu}
\end{equation}
and an effective bulk modulus
\begin{equation}\label{kappa}
\frac{1}{\kappa} = \frac{1}{K} -  \frac{1}{K +\frac{4}{3}\mu}.
\end{equation}
In the classical case 
\begin{equation}
\int d^4 x \longrightarrow \frac{1}{T} \int d^3 x
\end{equation}
and so we recover the classical effective action
\begin{equation}\label{cSeff}
S_{eff}[\psi ,\phi ] = 
\frac{1}{T}\int d^{3}x
\left[\frac{\kappa}{2}
\phi^{2}
+\frac{P^{2}}{2K}+ 
{\cal L}[\psi,b^{*} ] + \lambda (\phi + \frac{P}{K})\psi^{2}[\vec{x}] \right]
\end{equation}
with definitions as above.  
We note that in the main text 
we have replaced the renormalized $b^*$ by $b$ and we have set $P=0$ for 
presentational simplicity.

\subsection{Preliminaries}\label{}
The partition function can be written as an integral over the order
parameter and strain fields
\begin{equation}\label{}
Z = \int {\cal D}[\psi ]
\int {\cal D}[u]e^{- S[\psi ,u]}
\end{equation}
where $\psi $ is the order parameter field and $u (x)$ is the local
displacement of the lattice that determines the strain fields
according to the relation 
\begin{equation}\label{}
e_{ab} (x) = \frac{1}{2}\left(\frac{\partial u_{a}}{\partial x_{b}} +
\frac{\partial u_{b}}{\partial x_{a}}
\right).
\end{equation}
Here the action 
is
is determined by an integral over the 
Lagrangian $L$, $S= \int  d^{4}x L$.  In the quantum case, 
$\int d^{4}x \equiv \int d\tau \int d^{3}x$ is a space-time integral
over configurations that are periodic in 
imaginary time $\tau \in [ 0,\beta ]
$, where the $\beta$ inverse temperature ($k_B = 1$). 
In the classical case the
time-dependence disappears and the integral over $\tau $ is replaced
by $1/T$ so that $S = \frac{1}{T}\int d^{3}x L$. 

The action divides up into three parts 
\begin{equation}\label{action}
S = S_{A}+S_{I}+S_{B} = \int d^{4}x ( L_{A}[u]+ L_{I}[\psi,e]+L_{B}[\psi ]),
\end{equation}
where the contributions to the Lagrangian are: 
(i) a Gaussian term describing the elastic degrees of freedom in an
isotropic system
\begin{equation}\label{l1}
L_{A}[u] =
 \frac{1}{2}\left[ 
\rho 
\dot u_{l}^{2}+ \left(K-\frac{2}{3}\mu \right)
e_{ll}^{2}+ 2\mu e_{ab}^{2}\right]- \sigma_{ab}e_{ab} 
\end{equation}
where $\sigma_{ab}$ is the external stress and we have assumed a
summation convention in which repeated indices are summed over, so
that for instance,  $e_{ll} \equiv  \sum_{l=1,3}e_{ll}$
 and
$e_{ab}^{2}\equiv e_{ab}e_{ab}= \sum_{a,b=1,3}e_{ab}^{2}$
;
(ii) an interaction  term
\begin{equation}\label{l2}
L_{I}[\psi ,e]= \lambda e_{ll}\psi^{2}
\end{equation}
describing the coupling between the volumetric strain $e_{ll}= {\rm Tr}[e]$
and the ``energy density'' $\psi^{2}$ of the order parameter
$\psi$; (iii)  
the Lagrangian $L_{B}[\psi ]$ 
of the order parameter that, in the simplest case, is a $\psi^{4}$ field
theory
\begin{equation}\label{l3}
L_{B}[\psi,b] = 
\frac{1}{2} (\partial_{\mu}\psi
)^{2}+ \frac{a}{2}\psi^{2}+ \frac{b}{4!}\psi^{4},
\end{equation}
where we have explicitly noted its dependence on the interaction
strength $b$.
At a finite temperature critical point, 
all time-derivative terms are dropped from these expressions. 

Since the integral over the strain fields is Gaussian,  
the latter can be integrated out of the partition
function leading to an 
effective action of the $\psi $ fields 
$S_{eff}[\psi]= S_{B}[\psi ]+ \Delta S[\psi ]$ where
\begin{equation}\label{}
e^{-\Delta S[\psi ]}=  \int {\cal D}[u]e^{- (S_{A}+S_{I})}.
\end{equation}
If we write the elastic action in the schematic, discretized form 
\begin{equation}\label{}
S_{A}+S_{I}= \frac{1}{2} \sum_{i,j}
u_{i}M_{ij} u_{j} + \lambda \sum_{j}u_{j} \psi_{j}^{2}
\end{equation}
then the effective action becomes simply 
\begin{equation}\label{}
\Delta  S  = \frac{1}{2} \ln {\rm det}[M] - 
\frac{\lambda^{2}}{2}
\sum_{i,j} \psi_{i}^{2}M^{-1}_{i,j}\psi_j^2
\end{equation}
where the second term is recognized as an induced attractive interaction between
the order-parameter fields. 
The  subtlety in this procedure derives from the division of 
the strain field into two parts: a uniform contribution determined by
boundary conditions and a 
fluctuating component in the bulk. 
For the classical case
\begin{equation}\label{elabel}
e_{ab} (\vec{x}) = e_{ab} + \frac{1}{\sqrt{V}} \sum_{\vec{q}\neq 0} \frac{i}{2}
\left(q_{a}u_{b} (\vec{q}) + q_{b}u_{a} (\vec{q}) \right)e^{ i \vec{q}\cdot \vec{x}}
\end{equation}
where the $u_{b} (\vec{q})$ are the Fourier transform of the atomic
displacements, while in the quantum problem
\begin{equation}\label{}
e_{ab} (\vec{x},\tau ) = e_{ab} + \frac{1}{\sqrt{V\beta} } \sum_{i\nu_n}
\sum_{\vec{q}\neq 0} \frac{i}{2}
\left(q_{a}u_{b} (q) + q_{b}u_{a} (q) \right)e^{ i (\vec{q}\cdot \vec{x}-\nu_n\tau)},
\end{equation}
where $\nu_n=2\pi n T$ is the bosonic Matsubara frequency. Note that
the exclusion of all terms where $\vec{q}=0$ from the summation also
excludes the special point where both $i\nu_n$ and $\vec{q}$ are zero. As we now
demonstrate, the overall attractive interaction ($\propto -\psi_{i}^{2}M^{-1}_{ij}\psi_{j}^{2}$) contains both
short-range and infinite range components. 

\subsection{The Gaussian Strain integral: Classical Case }\label{}

Our task is to calculate the Gaussian integral,
\begin{equation}\label{}
e^{-\Delta S[\psi] }  = \int {\cal D}[e_{ab},u_{q}]e^{- (S_{A}+S_{I})}
\end{equation}
where the classical action
\begin{equation}\label{}
S_{A}+S_{I} = 
\frac{1}{T} \int d^{3}x \left[  \frac{1}{2}
\left(K-\frac{2}{3}\mu \right)
e_{ll}^{2} (\vec{x}) + \mu e_{ab} (\vec{x})^{2}
+ (\lambda \psi^{2} (\vec{x})+ {P}) e_{ll} (\vec{x}) \right],
\end{equation}
where we have denoted $\sigma_{ab}= -{P}\delta_{ab}$ in terms
of the pressure $P$. 
We begin by splitting the strain field into the $q=0$ and finite $q$
components, 
\begin{equation}
e_{ab} (\vec{x}) = e_{ab} + \frac{1}{\sqrt{V}} \sum_{\vec{q}\neq 0} \frac{i}{2}
\left(q_{a}u_{b} (\vec{q}) + q_{b}u_{a} (\vec{q}) \right)e^{ i \vec{q}\cdot \vec{x}}.\tag{\ref{elabel}}
\end{equation}
This separation enables us to use periodic boundary conditions,
putting the system onto a spatial torus with discrete momenta
$\vec{q}= \frac{2\pi}{L} (l,m,n)$.
After this transformation, the action divides up into two terms, 
$S= S[e_{ab},\psi ]+S[u,\psi ]$. 
We shall define the integrals
\begin{equation*}
\int de_{ab} e^{-S[e_{ab},\psi ]} = e^{-S_{1}[\psi ]},
\end{equation*}
and 
\begin{equation}\label{}
\int {\cal D}[u]e^{-S[u,\psi ]} = e^{-S_{2}[\psi]}.
\end{equation}

The uniform part of the action is
\begin{eqnarray}\label{l}
S[e_{ab},\psi ] &=& 
\frac{V}{T} \left[  \frac{1}{2}
\left(K-\frac{2}{3}\mu \right)
e_{ll}^{2}  +\mu e_{ab}^{2}\right]
+ \frac{V}{T} (\lambda \psi^{2}_{q=0}+P) e_{ll} \cr
&=& \frac{1}{2} e_{ab}{\cal M}_{abcd} e_{cd} + v_{ab}e_{ab},
\end{eqnarray}
where
$\psi^{2}_{\vec{q}}= \frac{1}{V}\int d^{3}x\psi^{2} (\vec{x})e^{i \vec{q}\cdot
\vec{x}}
$
 is the Fourier transform of the fluctuations in ``energy
density'' and
\begin{eqnarray}\label{curlym}
{\cal M}_{abcd} &=& K 
\overbrace {(\delta_{ab}\delta_{cd})}^{{\cal P}^{L}_{abcd}}+
2\mu
\overbrace {\left(\delta_{ac}\delta_{bd}-\frac{1}{3}\delta_{ab}\delta_{cd}
\right) }^{{\cal P}^{T}_{abcd}}, \\
v_{ab}&=& \frac{V}{T} (\lambda \psi^{2}_{q=0}+P)\delta_{ab}.
\end{eqnarray}
The nonuniform part of the action is  
\begin{equation}\label{}
S[u,\psi ] 
= 
 \frac{1}{T}
\sum_{\vec{q}\neq 0} \left( \frac{1}{2}
u^{*}_{a}({\vec{q}}) M_{ab}
u_{b} (\vec{q}) + \vec{a} (\vec{q})
\cdot \vec{u} (\vec{q})\right)
\end{equation}
where
\begin{eqnarray}\label{l}
M_{ab}&=& \left[ 
\left(K - \frac{2}{3}\mu \right)
q_{a}q_{b}
+ \mu \left(q^{2}\delta_{ab}+q_{a}q_{b} \right)
\right],\cr
\vec{a}_{q}&=& 
\left( 
 {i\lambda
}{\sqrt{V}}
\ \psi^{2}_{-q}\right) \vec{q}.
\end{eqnarray}

When we integrate over the uniform part of the strain field,
\begin{eqnarray}\label{uni1}
 \frac{1}{2} e_{ab}{\cal M}_{abcd} e_{cd} + v_{ab}e_{ab}\rightarrow 
S_{1}[\psi ]= - \frac{1}{2}v_{ab}{\cal M}^{-1}_{abcd}v_{cd}
\end{eqnarray}
Now   the two terms $P^{L}_{abcd}$ and $P^{T}_{abcd}$ in  ${\cal M}$ (\ref{curlym}) are independent
projection operators
($P^{\Gamma}_{abef}P^{\Gamma}_{efcd}=P^{\Gamma}_{abcd}$, $\Gamma\in L,
T$), projecting the longitudinal and transverse components of the
strain. The inverse of ${\cal M}$ is then given by 
\begin{equation}\label{}
{\cal M}^{-1}_{abcd} = \frac{T}{V}\left[\frac{1}{K} (\delta_{ab}\delta_{cd})+
\frac{1}{2\mu}\left(\delta_{ac}\delta_{bd}-\frac{1}{3}\delta_{ab}\delta_{cd}
\right) \right],
\end{equation}
so the Gaussian integral over the uniform part of the strain field
gives
\begin{equation}\label{result1}
S_{1}[\psi] = - \frac{1}{2}v_{ab}{\cal M}^{-1}_{abcd}v_{cd} =  
-\frac{V}{2T} \frac{1}{K} (\lambda\psi^{2}_{q=0}+P)^{2}.
\end{equation}

Now the matrix entering the fluctuating part of the action $S[u,\psi]$, can be 
projected into the longitudinal and transverse components of the strain 
\begin{equation}\label{}
M_{ab} (\vec{q}) = q^{2} \left[ 
\left(K  +\frac{4}{3}\mu \right)
\hat  q_{a}\hat q_{b}
+ \mu (\delta_{ab}- \hat  q_{a}\hat q_{b})
\right]
\end{equation}
where  $\hat q_{a}=q_{a}/q$ are the direction cosines of $\vec{q}$.  The inversion of this
matrix is then 
\begin{equation}\label{}
M^{-1}_{ab} (\vec{q}) = q^{-2} \left[ 
\left(K  +\frac{4}{3}\mu \right)^{-1}
\hat  q_{a}\hat q_{b}
+ \mu^{-1}(\delta_{ab}-\hat  q_{a}\hat q_{b})
\right],
\end{equation}
so the Gaussian integral over fluctuating part of the strain field
leads to
\begin{eqnarray}\label{l}
 \frac{1}{T}
\sum_{\vec{q}\neq 0}\frac{1}{2}
u^{*}_{a}({\vec{q}}) M_{ab} (\vec{q})
u_{b} (\vec{q}) + \vec{a} (\vec{q})
\cdot \vec{u} (\vec{q})&\rightarrow& \cr
S_{2}[\psi ]&=& -\frac{1}{2T}\sum_{\vec{q}\neq  0} a_{a} (-\vec{q} )
M^{-1}_{ab} (\vec{q}) a_{b} (\vec{q})\cr
&=&-\frac{V}{2T} \sum_{\vec{q}\neq  0 }  \psi^{2}_{-q}\psi^{2}_{q}\frac{\lambda^{2}}{K+\frac{4}{3}\mu}
\end{eqnarray}
We can rewrite this as a sum over {\sl all} $\vec{q}$, plus a 
remainder at $\vec{q}=0$:
\begin{eqnarray}\label{result2}
S_{2}[\psi]
&=& 
-\frac{V}{2T} \sum_{\vec{q}}
\psi^{2}_{-q}\psi^{2}_{q}\frac{\lambda^{2}}{K+\frac{4}{3}\mu}
+\frac{V}{2T} (
\psi^{2}_{q=0})^{2}\frac{\lambda^{2}}{K+\frac{4}{3}\mu}\cr
&=& 
- \frac{1}{2T}\frac{\lambda^{2}}{K+\frac{4}{3}\mu}\int d^{3}x\psi^{4}
(\vec{x})
+\frac{V}{2T} (
\psi^{2}_{q=0})^{2}\frac{\lambda^{2}}{K+\frac{4}{3}\mu}
.
\end{eqnarray}
The first term is a local attraction while the second term, involving
only the $\vec{q}=0$ Fourier component, corresponds to 
to a repulsive long range interaction. 

When we combine the results of the two integrals (\ref{result1}) and
(\ref{result2}) we obtain
\begin{equation}\label{annoy1}
\Delta S[\psi ] = 
-\frac{V}{2T}
 \frac{\lambda^{2}}{\kappa} (\psi^{2}_{q=0})^{2}
- \frac{1}{2T}\frac{\lambda^{2}}{K+\frac{4}{3}\mu}\int d^{3}x\psi^{4}
(x)
-\frac{V}{2T} \frac{1}{K} (2\lambda\psi^{2}_{q=0}P+P^{2})
\end{equation}
where 
\begin{equation}\label{}
\frac{1}{\kappa} = \frac{1}{K}- \frac{1}{K + \frac{4}{3}\mu}
\end{equation}
is the effective Bulk modulus.   

The final step in the procedure, is to carry out a Hubbard
Stratonovich transformation, factorizing the long-range
attraction in terms of a stochastic uniform field $\phi $,
\begin{equation}\label{annoy2}
-\frac{V}{2T}
 \frac{\lambda^{2}}{\kappa} (\psi^{2}_{q=0})^{2}
\rightarrow \frac{1}{T}\int d^{3}x\left[\frac{\kappa}{2}\phi^{2} +
\lambda \phi  \psi^{2} (x) \right].
\end{equation}
Combining (\ref{annoy1}) and (\ref{annoy2}) we obtain the following
expression for 
\begin{equation}\label{}
\Delta S[\psi ,\phi ] = \frac{1}{T}\int d^{3}x\left[
\frac{\kappa}{2}\phi^{2} + \frac{P^{2}}{2K} + \lambda\left(\phi
+\frac{P}{K} \right)\psi^{2} (x) - \frac{\lambda^{2}}{2 (K+\frac{4}{3}\mu)}\psi^{4} (x)
 \right].
\end{equation}
Finally, adding this term to the original order parameter action $S_{B}[\psi ]
= \frac{1}{T}\int d^{3}xL_{B}[\psi ,b]$, 
our final partition function can be written 
\begin{equation}\label{}
Z = \int d \phi \int {\cal D}[\psi] e^{-S_{eff}[\psi ,\phi ]}
\end{equation}
where $S_{eff}[\psi ,\phi ] = S_{B}[\psi ]+ \Delta S[\psi ,\phi ]$ is
given by
\begin{equation}\label{cSeff}
S_{eff}[\psi ,\phi ] = 
\frac{1}{T}\int d^{3}x
\left[\frac{\kappa}{2}
\phi^{2}
+\frac{P^{2}}{2K}+ 
{\cal L}[\psi,b^{*} ] + \lambda (\phi + \frac{P}{K})\psi^{2}[x] \right]
\end{equation}
where 
\[
{\cal L}[\psi ,b^{*}]= 
\frac{1}{2} (\partial_{\mu}\psi
)^{2}+ \frac{a}{2}\psi^{2}+ \frac{b^{*}}{4!}\psi^{4}.
\]
is the $\psi^{4}$ Lagrangian, with a renormalized short-range
interaction
\begin{equation}\label{rb}
b^{*}= b - \frac{12\lambda^{2}}{K+\frac{4}{3}\mu}.
\end{equation}
Note that in the main text we have dropped the ``$*$'' on $b$ for 
presentational simplicity; there $b$ refers to this renormalized interaction
(\ref{rb}).

Thus the main effects of integrating
out the strain field are a renormalization of the short-range
interaction of the order parameter field and the development of an
infinite-range interaction mediated by an effective strain field
$\phi $.   If we differentiate the action with respect to the
pressure, we  obtain the volumetric strain 
\begin{equation}\label{}
\frac{\delta S}{\delta P (\vec{x})} =  e_{ll} (\vec{x}) = \frac{1}{K}\left(P + \lambda \psi^{2} (\vec{x}) \right),
\end{equation}
which, as a result of integrating out the strain fluctuations, now
contains a contribution from the order parameter. Again in the main text we
set $P=0$ for presentational simplicity.

\subsection{The Gaussian Strain integral: Quantum Case }\label{}

In the quantum case, the action in the Gaussian strain integral
\begin{equation}\label{}
e^{-\Delta S[\Psi] }  = \int {\cal D}[e_{ab},u_{q}]e^{- (S_{A}+S_{I})}
\end{equation}
now involves an integral over space time, with 
$S = \int d^{4}x L\equiv \int_{0}^{\beta } d\tau
\int d^{3}x L$. We now restore the kinetic energy terms in Lagrangian (\ref{l1}) and
(\ref{l3}), so that now
the quantum action takes the form
\begin{equation}\label{}
S_{A}+S_{I} = 
\int d\tau d^{3}x \left[  
\frac{\rho }{2}\dot u_{l}^{2}+ \left(K-\frac{2}{3}\mu \right)
e_{ll}^{2} (x) + \frac{1}{2}2\mu e_{ab} (x)^{2}
+ (\lambda \psi^{2} (x)+ {P}) e_{ll} (x) \right].
\end{equation}
Again our task is to cast this
into matrix form
\begin{equation}\label{}
S_{A}+S_{I}= \frac{1}{2} \sum_{q}
u_{i}M_{ij} u_{j} + \lambda \sum_{j}u_{j} \psi_{j}^{2} \rightarrow  \frac{\lambda^{2}}{2}
\sum_{i,j} \psi_{i}^{2}M^{-1}_{i,j}\psi_j^2.
\end{equation}
where now the summations
run over the discrete
wavevector and Matsubara frequencies $q \equiv (i\nu_n,\vec q)$, where 
$\nu_n=\frac{2\pi }{\beta  }n$, $\vec{ q} = \frac{2\pi}{L} (j,l,k)$.
As before, we must separate out the static, $\vec{ q}=0$ component of
the strain tensor, writing 
\begin{equation}\label{}
e_{ab} (x,\tau ) = e_{ab} + \frac{1}{\sqrt{V\beta} } \sum_{i\nu_n}
\sum_{\vec{q}\neq 0} \frac{i}{2}
\left(q_{a}u_{b} (q) + q_{b}u_{b} (q) \right)e^{ i (\vec{q}\cdot \vec{x}-\nu_n\tau)},
\end{equation}
Note that there is no time-dependence to the uniform part of the
strain, since the boundary conditions are static. However the
fluctuating component excludes $\vec{q}=0$, but includes all Matsubara
frequencies; 
with these caveats,
the quantum integration of the strain fields closely follows that of the 
classical case.

Again 
the action divides up into two terms, 
$S= S[e_{ab},\psi ]+S[u,\psi ]$,
corresponding to the distinct unifom and finite $\vec{q}$ contributions
to the strain.
We shall again define the integrals
\begin{equation*}
\int de_{ab} e^{-S[e_{ab},\psi ]} = e^{-S_{1}[\psi ]},
\end{equation*}
and 
\begin{equation}\label{}
\int {\cal D}[u]e^{-S[u,\psi ]} = e^{-S_{2}[\psi]}.
\end{equation}

The uniform part of the action
\begin{eqnarray}\label{l}
S[e_{ab},\psi ] &=& 
\int d\tau  \left[  \frac{1}{2}
\left(K-\frac{2}{3}\mu \right)
e_{ll}^{2}  + \frac{1}{2}2\mu e_{ab}^{2}\right]
+ \frac{V}{T} (\lambda \psi^{2}_{q=0}+P) e_{ll} \cr
&=& \frac{1}{2} e_{ab}{\cal M}_{abcd} e_{cd} + v_{ab}e_{ab},
\end{eqnarray}
where 
\begin{eqnarray}\label{l}
{\cal M}_{abcd} &=& \left[K (\delta_{ab}\delta_{cd})+
2\mu\left(\delta_{ac}\delta_{bd}-\frac{1}{3}\delta_{ab}\delta_{cd}
\right) \right], \cr
v_{ab}&=& {V\beta } (\lambda \psi^{2}_{q=0}+P)\delta_{ab},
\end{eqnarray}
is unchanged, but now 
\begin{equation}\label{}
\psi^{2}_{q}= \frac{1}{V\beta } \int d^{4}x \psi^{2} (x)e^{-i (\vec{q}\cdot \vec{x}-\nu_n\tau)}
\end{equation}
is the space-time Fourier transform of the order parameter intensity.
The non-uniform part is now
\begin{equation}\label{}
S[u,\psi ] 
= 
 \sum_{i\nu_n}\sum_{\vec{q}\neq 0}
\left(
\frac{1}{2}
u^{*}_{a}(q) M_{ab}
u_{b} (q) + \vec{a} (q)
\cdot \vec{u} (q) \right),
\end{equation}
where 
\begin{eqnarray}\label{l}
M_{ab}&=& \left[ \rho \nu_n^2
\left(K - \frac{2}{3}\mu \right)
q_{a}q_{b}
+ \mu \left(q^{2}\delta_{ab}+q_{a}q_{b} \right)
\right],\cr
\vec{a}_{q}&=& 
\left( 
 {i\lambda
}{\sqrt{V\beta }}
\ \psi^{2}_{-q}\right) \vec{q}.
\end{eqnarray}

When we integrate over the uniform part of the strain field, we obtain
\begin{eqnarray}\label{l}
 \frac{1}{2} e_{ab}{\cal M}_{abcd} e_{cd} + v_{ab}e_{ab}&\rightarrow&\cr
 S_{1}[\psi ]&=& - \frac{1}{2}v_{ab}{\cal M}^{-1}_{abcd}v_{cd}\cr
&=&-\frac{V\beta }{2K}(\lambda\psi^{2}_{q=0}+P)^{2},
\end{eqnarray}
or 
\begin{equation}\label{}
S_{1}[\psi ]
=- \frac{1}{2K}\int d^{4}x (\lambda\psi^{2}_{q=0}+P)^{2}.
\end{equation}
For presentational simplicity, we will now set $P=0$ since the role of pressure 
here follows that in the classical treatment already described.

The matrix entering the fluctuating part of the action can be 
projected into the longitudinal and transverse components
\begin{equation}\label{}
M_{ab}= \left[ 
\left(\rho \nu_n^2 + (K  +\frac{4}{3}\mu ) \right)
\hat  q_{a}\hat q_{b}
+ \left(\rho \nu_n^2 + \mu \right)(\delta_{ab}- \hat  q_{a}\hat q_{b})
\right],
\end{equation}
where  $\hat q_{a}=q_{a}/q$ is the unit vector. The inversion of this
matrix is then 
\begin{equation}\label{}
M^{-1}_{ab}=  \left[
\frac{1}{\rho (\nu_n^2+c_L^2 q^{2})} 
\hat  q_{a}\hat q_{b}
+ \frac{1}{\rho (\nu_n^2+c_T^2q^2)}
(
\delta_{ab}-\hat  q_{a}\hat q_{b})
\right],
\end{equation}
where 
\begin{equation}\label{}
c_{L}^{2} = \frac{K+\frac{4}{3}\mu}{\rho }, \qquad c_{T}^{2} = \frac{2\mu}{\rho }
\end{equation}
are the longitudinal and transverse sound velocities.  The two terms
appearing in $M^{-1}$ are recognized as the propagators for
longitudinal and tranverse phonons.

When we integrate over the fluctuating component of the strain field,
only the longitudinal phonons couple to the order parameter:  
\begin{eqnarray}\label{l}
\frac{1}{2}
\sum_{i\nu_n}\sum_{\vec{q}\neq 0}
u^{*}_{a}(q) M_{ab} (q)
u_{b} (q) + \vec{a} (q)
\cdot \vec{u} (q)&\rightarrow& \cr
S_{2}[\psi ]&=& -\frac{1}{2}\sum_{i\nu_n}\sum_{\vec{q}\neq  0} a_{a} (-q )
M^{-1}_{ab} (q) a_{b} (q)\cr
&=&
-\frac{V\beta \lambda^{2}}{2 } \sum_{i\nu_n,\vec{q}\neq  0 }
\psi^{2}_{-q}\psi^{2}_{q}
\left(\frac{q^{2}}{\rho \nu_n^2+ (K+\frac{4}{3}\mu) q^2} \right).
\end{eqnarray}
Now in this last term, 
\begin{equation}\label{}
\left(\frac{q^{2}}{\rho \nu_n^2+ (K+\frac{4}{3}\mu) q^2} \right)
\end{equation}
the $\vec{q}=0$ term vanishes for any finite 
$\nu_n$, but in the case where $\nu_n=0$, the limiting
$\vec{q}\rightarrow 0$ form of this term is finite:
\begin{equation}\label{}
\left. 
\left(\frac{q^{2}}{\rho \nu_n^2+ (K+\frac{4}{3}\mu) q^2} \right)\right\vert_{\vec{q}\rightarrow 0} =
\left\{\begin{array}{cc}
0 & \nu_{n}\neq  0\cr
\frac{1}{K+\frac{4}{3}\mu} & \nu_n=0.
\end{array}
 \right.
\end{equation}
We can thus replace
\begin{equation}\label{l}
\sum_{i\nu_n,\vec{q}\neq  0 }
\psi^{2}_{-q}\psi^{2}_{q}
\left(\frac{q^{2}}{\rho \nu_n^2+ (K+\frac{4}{3}\mu) q^2}
\right)
\rightarrow 
\sum_{i\nu_n,\vec{q}}
\psi^{2}_{-q}\psi^{2}_{q}
\left(
\frac{q^{2}}{\rho \nu_n^2+ (K+\frac{4}{3}\mu) q^2} 
\right)
-
 \frac{(\psi^{2}_{q=0})^{2}}{K+\frac{4}{3}\mu}.
\end{equation}
so that 
\begin{equation}\label{}
S_{2}[\psi ] = 
\frac{V\beta \lambda^{2}}{2 (K+\frac{4}{3}\mu)} (\psi^{2}_{q=0})^{2} - 
\frac{V\beta \lambda^{2}}{2}\sum_{i\nu_n,\vec{q}}
\psi^{2}_{-q}\psi^{2}_{q}
\left(
\frac{q^{2}}{\rho \nu_n^2+ (K+\frac{4}{3}\mu) q^2} 
\right).
\end{equation}
If we now combine $S_{1}$ and $S_{2}$, we obtain 
\begin{equation}\label{}
S_{1}+S_{2} = \frac{V\beta \lambda^{2}}{2\kappa } (\psi^{2}_{q=0})^{2} 
- 
\frac{V\beta \lambda^{2}}{2}\sum_{q}
\psi^{2}_{-q}\psi^{2}_{q}
\left(
\frac{q^{2}}{\rho \nu_n^2+ (K+\frac{4}{3}\mu) q^2} 
\right).
\end{equation}
where
\begin{equation}\label{}
\frac{1}{\kappa} = \frac{1}{K}- \frac{1}{K+\frac{4}{3}\mu}
\end{equation}
is the effective Bulk modulus, as in the classical case.

Next we carry out a Hubbard-Stratonovich transformation, rewriting the
the long-range attraction in terms of a stochastic static and uniform
scalar field $\phi $ as follows
\begin{equation}\label{}
-\frac{V\beta }{2}
 \frac{\lambda^{2}}{\kappa} (\psi^{2}_{q=0})^{2}
\rightarrow 
\int d^{4}x\left[\frac{\kappa}{2}\phi^{2} +
\lambda \phi  \psi^{2} (x) \right].
\end{equation}
The remaining interaction term can be divided up into two parts as
follows
\begin{equation}\label{}
\sum_{q}
\psi^{2}_{-q}\psi^{2}_{q}
\left(
\frac{q^{2}}{\rho \nu_n^2+ (K+\frac{4}{3}\mu) q^2} 
\right) 
= \frac{1}{K+\frac{4}{3}\mu}\sum_{q}\psi^{2}_{-q}\psi^{2}_{q}\left[1-
\left(
\frac{\nu_n^{2}/c_{L}^{2}}
{q^{2}+\nu_n^2/c_{L}^{2}}
\right)
\right] .
\end{equation}
The first term inside the brackets is independent of momentum and frequency,
leading to a finite local attraction term that will act to renormalize
the $b$ term in the Lagrangian ${\cal L}_{\psi ,b}$ as in the
classical case.  The second term is a non-local and retarded
interaction.  Due to Lorentz invariance, simple power-counting shows that this 
term has the same scaling dimensionality
as a local repulsive term, and thus it will not modify the critical behavior
of the second-order phase transition. 

If we transform back into 
into space-time co-ordinates, then we obtain
\begin{equation}\label{}
S_{1}+S_{2}= \int d^{4}x\left[\frac{\kappa}{2}\phi^{2} +
\lambda \phi  \psi^{2} (x) 
- \frac{\lambda^{2}}{2 (K+\frac{4}{3}\mu)}\psi^{4} (x)\right] + S_{NL}
= S_{eff}[\psi,\phi] + S_{NL}
\end{equation}
where
\begin{equation}\label{}
S_{NL} = \frac{\lambda^{2}}{2V\beta (K+\frac{4}{3}\mu)} \int d^{4}x d^{4}x'
\partial_{\tau } (\psi^{2}) (x) V (x-x')\partial_{\tau } (\psi^{2}) (x')
\end{equation}
and
\begin{equation}\label{}
V (x)= \int \frac{d^{4}q}{(2\pi)^{4}}\left(\frac{c_{L}^{-2}}{|\vec{q}|^{2}+\nu_n^2/c_L^2} \right)e^{i (\vec{q}\cdot\vec{x}-i\nu_n\tau)}= \frac{1}{2\pi c_{L}^{2}}\frac{1}{( |\vec{x}|^{2}+ c_{L}^{2} \tau^{2})}
\end{equation}
is the non-local interaction mediated by the acoustic phonons. Then
the final quantum partition function
resulting from integrating out the strain fields in the quantum case 
can be written ($P \neq 0$)
\begin{equation}\label{}
Z = \int d \phi \int {\cal D}[\psi] e^{-S_{eff}[\psi ,\phi ]- S_{NL}}
\end{equation}
where 
\begin{equation}\label{qSeff}
S_{eff}[\psi,\phi] = \int d^{4}x
\left[\frac{\kappa}{2}\phi^{2}+ \frac{P^{2}}{2K}+ {\cal L}[\psi,b^{*} ] + \lambda \left(\phi+ \frac{P}{K}\right) \psi^{2}[x] \right]
\end{equation}
and
\[
{\cal L}[\psi,b^{*} ]= 
\frac{1}{2} (\partial_{\mu}\psi
)^{2}+ \frac{a}{2}\psi^{2}+ \frac{b^{*}}{4!}\psi^{4}.
\]
is the $\psi^{4}$ Lagrangian, with a renormalized short-range
interaction
\begin{equation}\label{bstar}
b^{*}= b - \frac{12\lambda^{2}}{K+\frac{4}{3}\mu}
\end{equation}
and an effective bulk modulus
\begin{equation}\label{}
\frac{1}{\kappa} = \frac{1}{K} -  \frac{1}{K +\frac{4}{3}\mu}.
\end{equation}
Thus, as in the classical case,  the main effect of integrating
out the strain field, is a renormalization of the short-range
interaction of the order parameter field, and the development of an
infinite range interaction, mediated by an effective strain field
$\phi $.  The introduction of a
nonlocal contribution with the same scaling dimensions as the $\psi^{4}$ term
will not affect the properties of the fixed point, and thus it will not
change the universality class of the fixed point, as in the classical
Larkin-Pikin case.  However we emphasize that in its quantum generalization
the effective dimension of the theory is $d_{eff} = d + z$.
Again we note that in the main text we have replaced the coefficient of the
renormalized interaction $b^*$ in (\ref{bstar}) by $b$ for presentational simplicity.

%\bibliography{QACP}
%\bibliographystyle{apsrev}
\end{document}